\documentclass{article}
\usepackage{amsmath,graphicx,mlspconf}
\usepackage{nonfloat}
\usepackage{graphicx}
\usepackage{amssymb}
\usepackage{booktabs}
\usepackage{epstopdf}
\usepackage{amsmath}
\usepackage{amsthm}
\usepackage{algorithm}
\usepackage{multirow}
\usepackage{algorithmic}

\toappear{2012 IEEE International Workshop on Machine Learning for Signal Processing, Sept.\ 23--26, 2012, Santander, Spain}

\twoauthors
  {Shang Shang, Sanjeev R. Kulkarni, Paul W. Cuff}
	{Department of Electrical Engineering\\
	Princeton University\\ Princeton, NJ, 08540, U.S.A.}
  {Pan Hui}
	{Deutsche Telekom
Laboratories\\
	Ernst-Reuter-Platz 7\\ 10587 Berlin, Germany}

\begin{document}
%
\title{ A RANDOM WALK BASED MODEL INCORPORATING SOCIAL INFORMATION FOR RECOMMENDATIONS}
\maketitle
\begin{abstract}
Collaborative filtering (CF) is one of the most popular approaches to build a recommendation system. In this paper, we propose a hybrid collaborative filtering model based on  a Makovian random walk to address the data sparsity and cold start problems in recommendation systems. More precisely, we construct a directed graph whose nodes consist of items and users, together with item content, user profile and social network information. We incorporate user's ratings into edge settings in the graph model. The model provides personalized recommendations and predictions to individuals and groups.  The proposed algorithms are evaluated on MovieLens and Epinions datasets. Experimental results show that the proposed methods perform well compared with other graph-based methods, especially in the cold start case.
\end{abstract}
\begin{keywords}
Recommendation system, random walk, social networks, hybrid collaborative filtering model
\end{keywords}
\section{Introduction}
\label{sec:intro}
Over the last decade, the commercialization of early generations of recommendation systems achieved great success. Recommendation systems serve as an important component of online retail and Video on Demand (VoD) services such as Amazon and Netflix \cite{Mnih}.  Recommenders typically provide the target user a list of customized recommendations through collaborative filtering or content-based filtering. Intensive work has been done to improve the performance of both of these techniques.  Traditional recommendation systems assume that users are independent, and recommendations are given according to users' explicit or implicit rating history and/or item content information\cite{Fouss}\cite{Su}. Problems such as \emph{data sparsity}, \emph{cold start}, and \emph{shilling attack} still challenge the design of recommendation systems \cite{Su}. User profile and social information, on the other side, provides extra information on user preference. This information is especially helpful in the case of giving recommendations to a new user with little or no rating history. The emergence of  e-commerce and online social networks provides us a good opportunity to integrate user social information into the recommendation model, so as to improve the recommendation results  or to alleviate the cold start problem \cite{He}\cite{Palau}. 

Collaborative filters use the known preferences of users to make recommendations or predictions to a target user. Memory-based collaborative filtering uses the entire user-item database to calculate the similarity value between users or items, and then a weighted sum is taken as a prediction for the target user on a certain item. See, for example, GroupLens \cite{Resnick}. Model-based approaches such as Bayesian Belief Net CF \cite{Su2} and regression-based CF \cite{Vucetic} learn a complex pattern from training data and use the model to predict a user's preference. The most related work are \cite{Fouss}\cite{Gori}\cite{Jamali}\cite{Bogers}. Fouss, et al. \cite{Fouss} suggested a dissimilarity measure between nodes of a graph, the expected commute time between two nodes, which the authors applied to collaborative filtering. Specially, they constructed an indirected bipartite graph where nodes are users and movies. A link is placed if the user watched that movie. Movies are then ranked in an ascendending order according to the average commute time to the target node. Gori et al \cite{Gori} built their graph model by only using items as nodes. In \cite{Gori}, two nodes are connected if at least one user rated both nodes. The weight of the edge is set as the number of users who rated both of the nodes. A random-walk based algorithm is then used to rank items according to the target user's preference record. In \cite{Jamali}, the authors combine the trust based and collaborative filtering approaches for recommendation. Target users take a finite-step random walk on a trust network, so as to use the ratings by trusted users to assist prediction. More recently,  Bogers \cite{Bogers} proposed ContextWalk, a collaborative filtering method to include different types on contextual information by taking random walks over the contextual graph. 

\begin{figure}[!t] 
\centering
  \includegraphics[width = .45\textwidth]{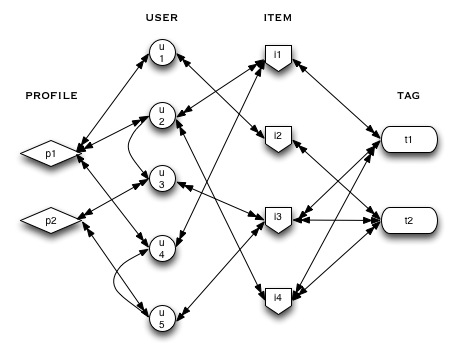}
  \caption{Hybrid collaborative filtering graph example.}
  \label{model}
\end{figure}

In this paper, we propose a random walk based hybrid collaborative filtering model that incorporates the social information of users. It is shown in \cite{Kleinberg} that a random walk approach is very effective in link prediction on social networks. Inspired by \cite{Kleinberg} and \cite{Page}, we create a recommendation graph, as shown in Fig. \ref{model}, consisting of items, users, item genres, and user profile information as nodes. Similar to PageRank, the stable distribution resulting from a random walk on the graph is interpreted as a ranking of the nodes for the purpose of recommendation and prediction. The structure of the collaborative filtering part of the recommendation graph is similar as the graph proposed in \cite{Fouss} and \cite{Bogers} in the means of connecting the user $u$ node and item node $i$ if there is a rating record of $u$ on $i$. Unfortunately, in \cite{Bogers}, the author did not provide experimental result to evaluate the performance, and the edge settings for constructing the network are not clear. In \cite{Fouss}, the authors assigned unit weight for the edges in the graph which cannot capture the user preference effectively. The expected commute time between item and user nodes was taken as the similarity measure to make recommendation. In \cite{Fouss} and \cite{Bogers}, the authors only gave a list of recommended items; no rating prediction is available. In this paper,  the edges of the graph is related to user rating score instead of  simply  being set to a unit value. Apart from the collaborative filtering graph which only contains user rating information, we add user social profile and social network information, which makes it possible to provide customized recommendation to new users even if no previous rating information is available.  
The main contribution of this paper is: (1) we propose a hybrid collaborative filtering graph model incorporating user social network, user profile information, together with item content and user-item rating history together to give recommendations; (2) we describe in detail the construction and edge weight assignment which reflect user preferences effectively; (3) we extend the application of the recommendation algorithm to group recommendation; (4) we design experiments on multiple data sets to evaluate the performance of proposed algorithm.  

In a typical setting, there is a list of $m$ users $\mathcal{U} = \{u_1, u_2, ..., u_m\}$ and a list of $n$ items $\mathcal{I} = \{i_1, i_2,..., i_n\}$. Each user $u_j$ has a list of items $I_{u_j}$, that the user has rated or from which the user's preference can be inferred. The ratings can either be explicit, for example, on a 1-5 scale as in Netflix, or implicit such as purchases or clicks. These data form a $m \times n$ rating matrix $R$, where $R_{ui}$ denotes the rating of user $u$ on item $i$. Assume that binary tagging and user social information is given. Let $\mathcal{T} = \{t_1, t_2, ..., t_k\}$ be the set of tagging information of items. For example, for movies, $\mathcal{T}$ can be genre, main actor, release date, etc.  $T_i \in \{0,1\}^k$ denotes the features of item $i$, where $k$ is the total number of tags. Correspondingly, let $\mathcal{P} = \{p_1, p_2, ..., p_l\}$ be the set of user profile information, including age, occupation, gender,  etc.. $P_u \in \{0,1\}^l$ denotes the profile features of user $u$, where $l$ is the dimension of the features of all users.   $\mathcal{S} = (\mathcal{U},\mathcal{E}_s)$ contains social network information, represented by an undirected or directed graph, where $\mathcal{U}$ is a set of nodes and $\mathcal{E}_s$ is a set of edges.  For all $ u, v \in \mathcal{U}$, $(u,v)\in \mathcal{E}$ if $v$ is a friend of $u$. We want to make recommendations for a target user or a group of users given the above information.

The rest of this paper is organized as follows. We propose our random walk based recommendation model in Section \ref{sec: individuals}. The performance of the proposed model is evaluated in Section \ref{sec:Ex}, followed by conclusions and acknowledgements in Section \ref{sec: conclusions} and \ref{sec: ack}.

\section{A Hybrid Collaborative Filtering Model Based on Random Walks}
\label{sec: individuals}
In this section, we will describe our algorithm in detail, inspired by Google's PageRank. Specially we describe how to construct the graph and make recommendations. 

PageRank \cite{Page} calculates a probability distribution representing the likelihood that a web surfer randomly clicking on web links will arrive at any webpage. A similar approach can be used for movie recommendation. Every time a user has watched a movie, the system may show some more movies that other users who like this one also like. As in PageRank, there is a damping factor to indicate that the movie watcher may finally stop browsing. Now the key issue is how to construct this recommendation graph and represent flow on the graph.

\subsection{Graph construction}

\subsubsection{Graph settings}
Let $\mathcal{G} = \{\mathcal{V}, \mathcal{E}\}$ be a directed graph model for CF, where $\mathcal{V}:= \mathcal{U}\cup \mathcal{I} \cup \mathcal{T} \cup \mathcal{P}$. The nodes of the graph consist of users, items, item information and user profiles. For $v_i, v_j \in \mathcal{V}$, $(v_i, v_j)\in \mathcal{E}$ if and only if there is an edge point from $v_i$ to $v_j$, which is determined as given below. The weight are specified in the next subsection.
\begin{itemize}
\item For $u\in \mathcal{U}, i \in {\mathcal{I}}$, $(u, i) \in \mathcal{E}$ and $(i, u) \in \mathcal{E}$ if and only if $R_{ui} \ne 0$, i.e., an item $i$ and a user $u$ are connected if there is a rating records of user $u$ on item $i$, with weight $w_{ui}$ and $w_{iu}$. 
\item For $i \in \mathcal{I}, t\in \mathcal{T}$, $(i, t)\in \mathcal{E}$ and $(t, i)\in \mathcal{E}$ if and only if $T_{i}^{(t)} \ne 0$, i.e., the item $i$ and tag $t$ are connected if $i$ is tagged by $t$, with weight $w_{it}$ and $w_{ti}$.
\item For $u\in \mathcal{U}, p\in {\mathcal{P}}$, $(u, p) \in \mathcal{E}$ and $(p, u) \in \mathcal{E}$ if and only if $P_{u}^{(p)} \ne 0$, i.e., a profile feature $p$ and a user $u$ are connected if the user $u$ belongs to the profile category $p$, with weight $w_{up}$ and $w_{pu}$. 
\item For $u_1, u_2 \in \mathcal{U}$, $(u_1, u_2)\in \mathcal{E}$ if and only if $(u_1, u_2)\in \mathcal{E}_s$, with weight $w_{u_1u_2}$. Note that the relationship in social networks is not necessarily mutual, it could be a unilateral relationship such as in Twitter\footnote{twitter.com}, epinions\footnote{www.epinions.com}, etc.
\end{itemize}

\subsubsection{Edge weight assignment}
\label{sec:weight}
The main part of our rank graph is the collaborative filtering graph, which includes the user nodes, item nodes and the edges between them. The weights of edges in the collaborative filtering graph can be assigned as follows:
\begin{equation}
w_{ui} = w_{iu} = \exp\left({\frac{r_{ui} - \bar{r}_u}{\sqrt{\sum_{i\in I_{u}}(r_{ui} - \bar{r}_u)^2}}}\right),
\end{equation}
\begin{equation}
\bar{r}_u := \frac{\sum_{i\in I_u}r_{ui}}{|I_u|}.
\end{equation}
where $I_{u}$ denote the set of items which user $u$ has rated. Note that a larger edge weight indicates more chance that the random walk passes through that edge. If user $u$'s rating on item $i$ $r_{ui}$ is lower than the average rating $\bar{r}_u$, $w_{ui}$ and $w_{iu}$ are less than 1; otherwise are greater than 1. The assignment of weights do not depend on the variance of the user's ratings.


For the extended graph, i.e. nodes and edges containing item content, user profile or social network information, we simply assign an edge weight of 1 if an edge is present.

\subsection{Rank score computation}
 \subsubsection{Random walk on a weighted graph}
 A random walk is a Markov process with random variables $X_1, X_2, ..., X_t, ...$ such that the next state only depends on the current state. For a random walk on a weighted graph, $X_{t+1}$ is a vertex chosen according to the following probability distribution:
 \begin{equation}
P_{ij} := P(X_{t+1} = j|X_t = i) = \frac{w_{ij}}{\sum_{j \in \mathcal{N}_i}w_{ij}},
\end{equation}
where $\mathcal{N}_i$ are the neighbors of $i$,  
$\mathcal{N}_i := \{j  | (i,j) \in \mathcal{E}\}. $ 
As mentioned in Section \ref{sec:weight}, a higher weight indicates a higher chance that the random walk moves through that edge.

 \subsubsection{Rank score computation}
 \label{sec: RSC}
For the recommendation graph $\mathcal{G} = \{\mathcal{V}, \mathcal{E}\}$. Let $v = |\mathcal{V}|$ denote the number of nodes on the graph.  $m$ is a $v\times 1$ \emph{customerized probability vector}. 
\begin{equation}
\theta = e_u,
\end{equation}
where $e_1, e_2, ..., e_v$ are the standard basis of column vectors. $\beta$ is a \emph{damping factor}. With probability $1-\beta$, the random walk is teleported back to node $u$. The rank score $s$ satisfies the following equation:
\begin{equation}
 s = \beta Ws + (1- \beta)\theta,
 \label{pagerank}
 \end{equation}
  where $W$ is the weighted transition matrix with $W_{ij} = P_{ji}$. 
 
 So we have,
 \begin{equation}
s = \big(\beta W + (1 - \beta)\theta\textbf{1}^T\big)s := Ms
 \end{equation}
Hence the rank score is the \emph{principal eigenvector} of $M$, which can be computed by iterations fast and easily as shown below:
\begin{algorithmic}
\STATE $s^{(0)}_i  \leftarrow \frac{1}{v}$ for all $i$ 
\STATE $t = 1$ 
\WHILE{ $|s^{(t)} - s^{(t-1)}| < \epsilon$} 
	\FOR {$i = 1$ to $v$}
	\item
	$s^{(t)}_i = \sum_{j = 1}^v \beta W_{ij}s^{(t-1)}_i + (1 - \beta)\theta_i$
	\ENDFOR
	\STATE $t \leftarrow t + 1$
\ENDWHILE
\end{algorithmic}

Similar to PageRank, the rank score $s$ is interpreted as the importance of other nodes to the target user $u$. It is easy to see that we can increase the rank score by shortening the distance, adding more paths, or increasing the weight on the path to $u$. These are desired properties in a recommendation system. For example, even if item $i$ is not directly connected with $u$, but it is in the category to which many of $u$'s highly rated items belong, $i$ is very likely to have a high rank score. Or  if both user $u$ and $u'$ have similar opinions on a variety of  items, $u'$ will have high rank, so we can use $u'$'s explicit ratings to make recommendations and predictions for $u$.
\subsection{Recommendation}
\subsubsection{Direct method}
\label{direct}
Solving Equation (\ref{pagerank}) iteratively, we have a rank score of all nodes of the recommendation graph $\mathcal{G}$. Since the rank score represents the importance to the target user, we then separate and sort them according to the categories, i.e. users $\mathcal{U}$, items $\mathcal{I}$, tags $\mathcal{T}$ etc.  Sorted items excluding $I_u$ form a recommendation list to the target user $u$. We can compute the recommendation for every user.

\subsubsection{User-based recommendation}
\label{sec: user}
Similar to memory-based collaborative filtering which uses \emph{Pearson correlation} \cite{Resnick} as a similarity measure between users and items, we use rank score as an influence measure to make predictions. Given the rank score of the user set $\mathcal{U}$, we take the weighted sum of users' ratings on item $i$ as a prediction for the target user $u$,  as shown in Equation (\ref{user}):
\begin{equation}
\hat{r}_{ui}^{user} = \frac{\sum_{x\in U_i}s_x(r_{xi} - \bar{r}_x)}{\sum_{x\in U_i}s_x} + \bar{r}_u.
\label{user}
\end{equation} 
$s_x$ is the target user's personalized rank score of user $x$.  

\subsubsection{Item-based recommendation}
As in Section \ref{sec: user}, in order to perform an item-based recommendation, we can use the rank score of item set $\mathcal{I}$ as weight to predict the rating of item $i$ for the target user $u$.  As shown in Equation (\ref{item})
\begin{equation}
\hat{r}_{ui}^{item} = \frac{\sum_{j\in I_u}s_jr_{uj}}{\sum_{j\in I_u}s_j}.
\label{item}
\end{equation} 
In Equation (\ref{item}) we use $u$'s rating on similar items to predict the rating on $i$. $s_j$ is the target user's personalized rank score of item $j$. 
\subsection{Incremental computation}
In practice, the rating information and user's social information evolves. The recommendation graph changes when a new rating record is input, a new item is on sale, a new user registers, or even when a user changes his profile. Thanks to the popularity of PageRank, incremental computation of PageRank has been studied intensively \cite{Chen}\cite{Bahmani}. It is shown in \cite{Bahmani} that with a reset probability of $\epsilon$, the total work needed to maintain an accurate estimate of the PageRank of every node at all times is $O(\frac{n\ln m}{\epsilon ^2})$ in a network with $n$ nodes, and $m$  edges. Since it is beyond the scope of this paper, we do not address technical details for this problem.

\subsection{Discussions}
\subsubsection{Recommendations for groups}
Because of the special structure of the rank graph, we can naturally extend the recommendation for individual users to groups. Note that in order to give recommendations for individuals, we set the \emph{personalized vector} in Section \ref{sec: RSC} as $\textbf{e}_u$. Similarly, for recommendation for a group of users $\hat{u}$, we can set the \emph{personal vector} $\theta$ as
\begin{equation}
\theta = \frac{1}{|\hat{u}|}\sum_{u\in \hat{u}}e_u.
\end{equation} 
The rest of the predictions are same as described in the previous sections.
\subsubsection{Dealing with ``cold" users and ``cold" items}
A great challenge to recommendation systems resulting from data sparsity is the \emph{cold start} problem, namely, the question of how to effectively give recommendations to new users. A na\"{\i}ve approach is to provide the same recommendation to everyone. Studies show that two persons connected via a social relationship tend to have similar tastes, which is known as the ``homophily principle'' \cite{McPherson}. The availability of online social network offers us extra information about new users. Given the social network information, if a new user is connected with other nodes in our recommendation graph, we can then make personalized recommendation for the ``cold" user even if we do not have any rating information from this user.  Similarly, for ``cold" items we connect a new item in the recommendation graph according to its tagging information, so that we can then recommend the ``cold" item to users.  Experimental results are shown in Section \ref{sec:Ex}.


\section{Experiments}
\label{sec:Ex}
\subsection{Data sets}
\begin{figure}[!t] 
\centering
  \includegraphics[width = 3 in]{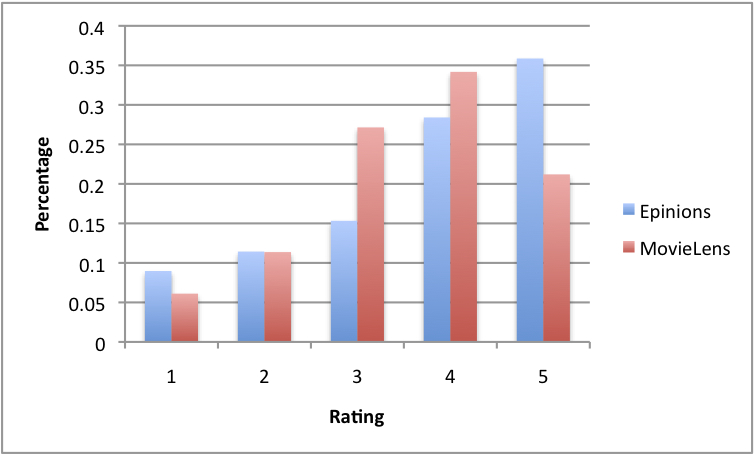}
  \caption{User rating distribution of Epinions and MovieLens datasets.}
  \label{fig: rating}
\end{figure}

\begin{figure}[!t] 
\centering
  \includegraphics[width = 3 in]{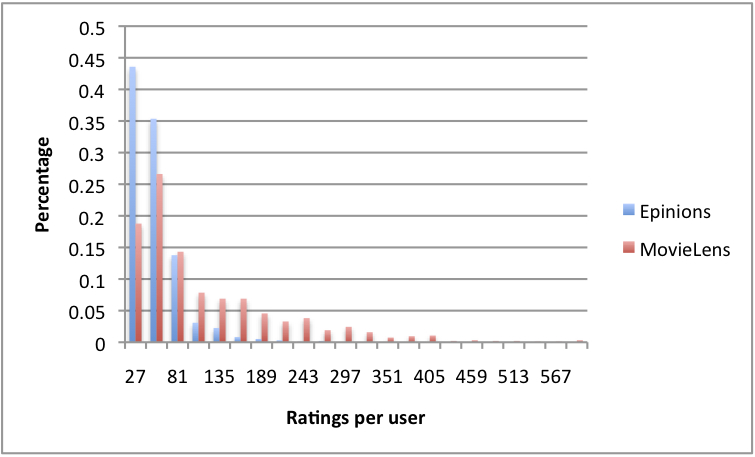}
  \caption{Histogram of ratings per user of Epinions and MovieLens datasets.}
  \label{fig: hist}
\end{figure}

In order to evaluate the performance of the proposed algorithm, we run experiments on Epinions and MovieLens data sets, both of which are widely used benchmarks for recommendation systems. Epinions is a website where users can post their reviews and ratings (1-5) on a variety of items (songs, softwares, TVs, etc.), as long as user's  \emph{web of trust}, i.e. ``reviewers whose reviews and ratings they have consistently found to be valuable" \cite{Massa}. We randomly select 946 items, 973 users and their trust network from Epinion data sets \cite{Massa} to perform the experiments.  The MovieLens data sets consists of 1682 movies and 943 users. Movies are labeled by 19 genres. User profile information such as age, gender and occupation is also available. User rating distributions and histograms of ratings per user for both data sets are shown in Fig. \ref{fig: rating} and Fig. \ref{fig: hist}.

\subsection{Experimental methodology and results}
We evaluate our results with two popular evaluation metrics for top-$k$ recommendations: recall and percentile. 

\emph{Recall}: In the top-$k$ recommendations, we consider any item in the top-$k$ recommendations that match any items in the testing set as a ``hit", as in \cite{Tso}. 
\begin{equation}
recall(k) = \frac{\#\textrm{hits of top-k}}{T},
\end{equation} 
where $T$ is the size of testing set. A higher recall value indicates a better prediction.

\emph{Percentile}: The individual percentile score is simply the average position (in percentage) that the item in the test set occupies in the recommendation list. For example, if four items are ranked 1st, 9th, 10th and 20th in a recommendation list consisting of 100 items, the percentile score is 0.1. A lower percentile indicates a better prediction.

In this experiment,  the test set $T$ contains all the 5-star rating records, thus we can consider them as relevant items for recommendation. The recommendation list has a length of 500 items for Epinions data sets and 900 for MovieLens data sets. We compare our methods UserRank CF (without social information) and UserRank in Section \ref{sec: individuals} with two state-of-art collaborative filtering methods L+ \cite{Fouss} and ItemRank \cite{Gori} described in Section \ref{sec:intro}. 

\begin{figure}[!h] 
\centering
  \includegraphics[width = 3 in]{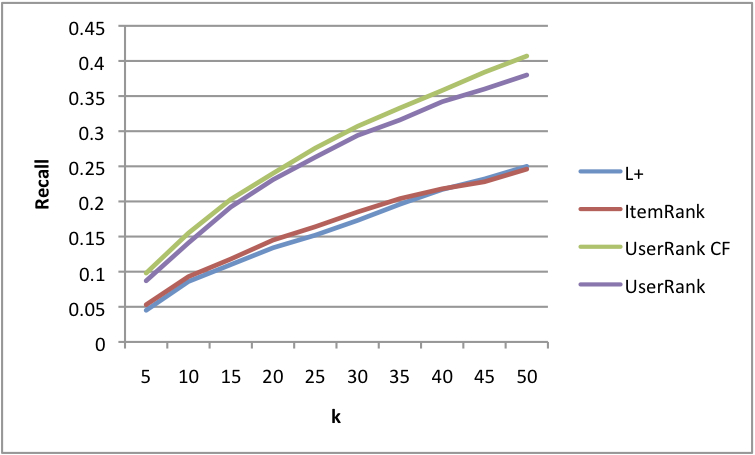}
  \caption{Epinion data sets top-k recall.}
  \label{erecall}
\end{figure}

\begin{figure}[!h] 
\centering
  \includegraphics[width = 3 in]{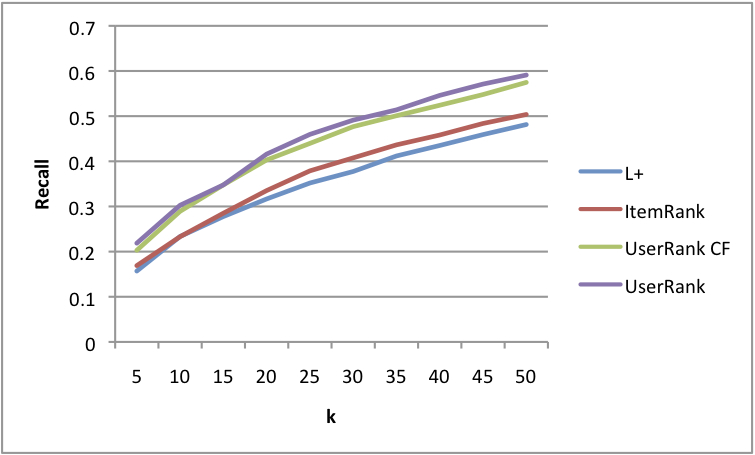}
  \caption{MovieLens data sets top-k recall.}
  \label{mrecall}
\end{figure}

 Experimental results of recall score are shown in Fig. \ref{erecall} and Fig. \ref{mrecall}. We can see that UserRank has a higher recall score in both data sets compared with baseline methods. However, in a ``warm start" scenario, adding social information does not change the performance much. In Table 1 and Table 2, we compared the percentile value for both warm start and cold start cases. It is worth noting that social information improves the performance of UserRank considerably in ``cold start" case.  

%


\begin{table}[htdp]
\begin{center}
\caption{Average percentile results obtained by 5-fold cross-validation for warm-start recommendation.} 
\begin{tabular}{lcc}

\hline
Methods & Epionions & MovieLens\\
\hline
L+ &  0.4192 & 0.1157\\
ItemRank &  0.3983 & 0.1150\\
UserRank CF   &  \bf{0.2325} & 0.081      \\
UserRank  with  social info.  & 0.2457  &     \bf{0.079} \\
\hline
\end{tabular}
\end{center}
\label{warm}
\end{table}

\begin{table}[htdp]
\begin{center}
\caption{Average percentile results obtained by 5-fold cross-validation for cold-start recommendation.} 
\begin{tabular}{lcc}

\hline
Methods & Epionions & MovieLens\\
\hline
ItemRank &  0.4592 & 0.1356\\
UserRank CF   &  0.3231 & 0.1204      \\
UserRank  with  social info.  & \bf{0.2874}  &     \bf{0.1112} \\
\hline
\end{tabular}
\end{center}
\label{cold}
\end{table}

\section{Conclusions}
\label{sec: conclusions}
In this paper, we present a hybrid collaborative filtering model based on a random walk for recommendation systems. It incorporates item content and user social information to make recommendations and predictions for target users. Social information improves the ``cold start" performance when lacking user rating information. Experiments are performed on two standard real-world data sets. The experimental results shows that the proposed method performs well compared to other state-of-art collaborative filtering methods.  

\section{Acknowledgements}
\label{sec: ack}
This research was supported in part by the Center for Science of
Information (CSoI), a National Science Foundation (NSF) Science and Technology Center, under grant
agreement CCF-0939370, by NSF under the grant CCF-1116013,
by the U.S. Army Research Office under grant number W911NF-07-1-0185, and by a research grant from Deutsche Telekom AG.

The authors would like to thank Dr. Guanchun Wang for valuable comments, and Mr. Julien Barbot for assistance in running the experiments.

\bibliographystyle{IEEEbib}
\bibliography{myrefs}

\end{document}